\documentclass[pre, twocolumn, amsmath,superscriptaddress, floatfix, byrevtev]{revtex4-1}
\usepackage{times}
\usepackage{mathptmx}
\usepackage[dvips]{graphicx}

\begin{document}
\title{Magnetized Fast Isochoric Laser Heating \\for Efficient Creation of Ultra-High-Energy-Density States}

\author{Shohei Sakata}
\affiliation{Institute of Laser Engineering, Osaka University, 2-6 Yamada-Oka, Suita, Osaka, 565-0871 Japan.}
\author{Seungho Lee}
\affiliation{Institute of Laser Engineering, Osaka University, 2-6 Yamada-Oka, Suita, Osaka, 565-0871 Japan.}
\author{Tomoyuki Johzaki}
\affiliation{Department of Mechanical Systems Engineering, Hiroshima University, Higashi-Hiroshima, Hiroshima, 739-8527, Japan.}
\author{Hiroshi Sawada}
\affiliation{Institute of Laser Engineering, Osaka University, 2-6 Yamada-Oka, Suita, Osaka, 565-0871 Japan.}
\affiliation{Department of Physics, University of Nevada Reno, Reno, Nevada 98557, USA.}
\author{Yuki Iwasa}
\affiliation{Institute of Laser Engineering, Osaka University, 2-6 Yamada-Oka, Suita, Osaka, 565-0871 Japan.}
\author{Hiroki Morita}
\affiliation{Institute of Laser Engineering, Osaka University, 2-6 Yamada-Oka, Suita, Osaka, 565-0871 Japan.}
\author{Kazuki Matsuo}
\affiliation{Institute of Laser Engineering, Osaka University, 2-6 Yamada-Oka, Suita, Osaka, 565-0871 Japan.}
\author{King Fai Farley Law}
\affiliation{Institute of Laser Engineering, Osaka University, 2-6 Yamada-Oka, Suita, Osaka, 565-0871 Japan.}
\author{Akira Yao}
\affiliation{Institute of Laser Engineering, Osaka University, 2-6 Yamada-Oka, Suita, Osaka, 565-0871 Japan.}
\author{Masayasu Hata}
\affiliation{Institute of Laser Engineering, Osaka University, 2-6 Yamada-Oka, Suita, Osaka, 565-0871 Japan.}
\author{Atsushi Sunahara}
\affiliation{Institute for Laser Technology, 1-8-4 Utsubo-honmachi, Nishi-ku Osaka, Osaka, 550-0004, Japan.}
\author{Sadaoki Kojima}
\affiliation{Institute of Laser Engineering, Osaka University, 2-6 Yamada-Oka, Suita, Osaka, 565-0871 Japan.}
\author{Yuki Abe}
\affiliation{Institute of Laser Engineering, Osaka University, 2-6 Yamada-Oka, Suita, Osaka, 565-0871 Japan.}
\author{Hidetaka Kishimoto}
\affiliation{Institute of Laser Engineering, Osaka University, 2-6 Yamada-Oka, Suita, Osaka, 565-0871 Japan.}
\author{Aneez Syuhada}
\affiliation{Institute of Laser Engineering, Osaka University, 2-6 Yamada-Oka, Suita, Osaka, 565-0871 Japan.}
\author{Takashi Shiroto}
\affiliation{Department of Aerospace Engineering, Tohoku University, Sendai, Miyagi 980-8579, Japan.}
\author{Alessio Morace}
\affiliation{Institute of Laser Engineering, Osaka University, 2-6 Yamada-Oka, Suita, Osaka, 565-0871 Japan.}
\author{Akifumi Yogo}
\affiliation{Institute of Laser Engineering, Osaka University, 2-6 Yamada-Oka, Suita, Osaka, 565-0871 Japan.}
\author{Alessio Morace}
\affiliation{Institute of Laser Engineering, Osaka University, 2-6 Yamada-Oka, Suita, Osaka, 565-0871 Japan.}
\author{Natsumi Iwata}
\affiliation{Institute of Laser Engineering, Osaka University, 2-6 Yamada-Oka, Suita, Osaka, 565-0871 Japan.}
\author{Mitsuo Nakai}
\affiliation{Institute of Laser Engineering, Osaka University, 2-6 Yamada-Oka, Suita, Osaka, 565-0871 Japan.}
\author{Hitoshi Sakagami}
\affiliation{National Institute for Fusion Science, National Institutes of Natural Sciences, 322-6 Oroshi, Toki, Gifu, 509-5292, Japan.}
\author{Tetsuo Ozaki}
\affiliation{National Institute for Fusion Science, National Institutes of Natural Sciences, 322-6 Oroshi, Toki, Gifu, 509-5292, Japan.}
\author{Kohei Yamanoi}
\affiliation{Institute of Laser Engineering, Osaka University, 2-6 Yamada-Oka, Suita, Osaka, 565-0871 Japan.}
\author{Takayoshi Norimatsu}
\affiliation{Institute of Laser Engineering, Osaka University, 2-6 Yamada-Oka, Suita, Osaka, 565-0871 Japan.}
\author{Yoshiki Nakata}
\affiliation{Institute of Laser Engineering, Osaka University, 2-6 Yamada-Oka, Suita, Osaka, 565-0871 Japan.}
\author{Shigeki Tokita}
\affiliation{Institute of Laser Engineering, Osaka University, 2-6 Yamada-Oka, Suita, Osaka, 565-0871 Japan.}
\author{Noriaki Miyanaga}
\affiliation{Institute of Laser Engineering, Osaka University, 2-6 Yamada-Oka, Suita, Osaka, 565-0871 Japan.}
\author{Junji Kawanaka}
\affiliation{Institute of Laser Engineering, Osaka University, 2-6 Yamada-Oka, Suita, Osaka, 565-0871 Japan.}
\author{Hiroyuki Shiraga}
\affiliation{Institute of Laser Engineering, Osaka University, 2-6 Yamada-Oka, Suita, Osaka, 565-0871 Japan.}
\author{Kunioki Mima}
\affiliation{Institute of Laser Engineering, Osaka University, 2-6 Yamada-Oka, Suita, Osaka, 565-0871 Japan.}
\affiliation{The Graduate School for the Creation of New Photonics Industries, 1955-1, Kurematsu, Nishi-ku, Hamamatsu, Shizuoka 431-1202, Japan.}
\author{Hiroaki Nishimura}
\affiliation{Institute of Laser Engineering, Osaka University, 2-6 Yamada-Oka, Suita, Osaka, 565-0871 Japan.}
\author{Mathieu Bailly-Grandvaux}
\affiliation{University of Bordeaux, CNRS, CEA, CELIA (Centre Lasers Intenses et Applications), UMR 5107, F-33405 Talence, France.}
\author{Jo$\tilde{a}$o Jorge Santos}
\affiliation{University of Bordeaux, CNRS, CEA, CELIA (Centre Lasers Intenses et Applications), UMR 5107, F-33405 Talence, France.}
\author{Hideo Nagatomo}
\affiliation{Institute of Laser Engineering, Osaka University, 2-6 Yamada-Oka, Suita, Osaka, 565-0871 Japan.}
\author{Hiroshi Azechi}
\affiliation{Institute of Laser Engineering, Osaka University, 2-6 Yamada-Oka, Suita, Osaka, 565-0871 Japan.}
\author{Ryosuke Kodama}
\affiliation{Institute of Laser Engineering, Osaka University, 2-6 Yamada-Oka, Suita, Osaka, 565-0871 Japan.}
\author{Yasunobu Arikawa}
\affiliation{Institute of Laser Engineering, Osaka University, 2-6 Yamada-Oka, Suita, Osaka, 565-0871 Japan.}
\author{Yasuhiko Sentoku}
\affiliation{Institute of Laser Engineering, Osaka University, 2-6 Yamada-Oka, Suita, Osaka, 565-0871 Japan.}
\author{Shinsuke Fujioka}
\email{sfujioka@ile.osaka-u.ac.jp}
\affiliation{Institute of Laser Engineering, Osaka University, 2-6 Yamada-Oka, Suita, Osaka, 565-0871 Japan.}

\begin{abstract}
The quest for the inertial confinement fusion (ICF) ignition is a grand challenge, as exemplified by extraordinary large laser facilities.
Fast isochoric heating of a pre-compressed plasma core with a high-intensity short-pulse laser is an attractive and alternative approach to create ultra-high-energy-density states like those found in ICF ignition sparks.
This avoids the ignition quench caused by the hot spark mixing with the surrounding cold fuel, which is the crucial problem of the currently pursued ignition scheme.
High-intensity lasers efficiently produce relativistic electron beams (REB). 
A part of the REB kinetic energy is deposited in the core, and then the heated region becomes the hot spark to trigger the ignition.
However, only a small portion of the REB collides with the core because of its large divergence.
Here we have demonstrated enhanced laser-to-core energy coupling with the magnetized fast isochoric heating.
The method employs a kilo-tesla-level magnetic field that is applied to the transport region from the REB generation point to the core which results in guiding the REB along the magnetic field lines to the core.
7.7 $\pm$ 1.3 \% of the maximum coupling was achieved even with a relatively small radial area density core ($\rho R$ $\sim$ 0.1 g/cm$^2$).
The guided REB transport was clearly visualized in a pre-compressed core by using Cu-$K_\alpha$ imaging technique. 
A simplified model coupled with the comprehensive diagnostics yields 6.2\% of the coupling that agrees fairly with the measured coupling.
This model also reveals that an ignition-scale areal density core ($\rho R$ $\sim$ 0.4 g/cm$^2$) leads to much higher laser-to-core coupling ($>$ 15\%), this is much higher than that achieved by the current scheme.
\end{abstract}

\maketitle


\section{Introduction}
A large energy, high power, high intensity laser can create a large volume ultra-high-energy-density plasma for particle accelerators \cite{Esarey2009}, planetary science \cite{Ohno2014,Smith2014}, astrophysics \cite{Remington2006,Fujioka2009a}, and nuclear physics \cite{Casey2017,Zylstra2016}.
Inertial confinement fusion (ICF) is an ultimate application of such extreme plasmas \cite{Lindl2014,Craxton2015,Betti2016}. 

The quest for the ICF ignition is a grand challenge, as exemplified by extraordinary large laser facilities.
A few mm-scale spherical capsule, which contains deuterium-tritium (DT) fusion fuel ice layer, is used in the ICF.
In the laser indirect-drive approach, the capsule surface is irradiated by X rays in a high-Z metal enclosure (hohlraum) driving a sequence of converging shock waves, and then the fusion fuel is compressed more than 1000 times solid density. 
Adiabatic compression heats up a DT gas initially filling the capsule interior, which then becomes the ignition spark at the final stage of the compression.

National Ignition Facility (NIF) and Laser MegaJoule (LMJ) have been constructed to demonstrate controlled thermonuclear fusion ignition by using the central ignition ICF scheme.
The scientific break-even, energy released by fusion reaction exceeds energy contains in the compressed fusion fuel, was achieved on NIF \cite{Hurricane2014}, however the pathway to the ICF ignition is still unclear.
One of the crucial problems of the current central ignition scheme is the hot spark mixing with the cold dense fuel because of significant growth of hydrodynamic instabilities during the compression.

Fast isochoric heating, also known as fast ignition \cite{Tabak1994}, of a pre-compressed core, was proposed as an alternative approach to the ICF ignition that avoids the ignition quench caused by the mixing because the hot spark is generated not by the adiabatic compression but by the external energy injection whose time scale is shorter than the hydrodynamic time scale ($<$ 100 ps).
Relativistic intensity laser pulses ($>$ 1.37 $\times$ 10$^{18}$ W $\mu$m$^2$ /cm$^2$) efficiently produce relativistic electron beams (REB) by laser-plasma interactions \cite{Wilks1992,Beg1997}.
The REB travels in a plasma from the REB generation point to the core.
A part of the REB kinetic energy is deposited in the core, and then the heated region becomes the hot spark to trigger the fusion ignition.

Several breakthroughs were described in past articles on fast heating research: the invention of the cone-in-shell target \cite{Kodama2001}, high area density core formation with the cone-in-shell target \cite{Theobald2014}, efficient laser-to-core coupling after reduction of preplasma filling in the cone \cite{Jarrott2016}.
Our achievement, namely enhanced laser-to-core energy coupling with the magnetized fast isochoric heating enabled by an application of external kilo-tesla-level magnetic field, is also significantly important, as a means to reduce the large radial spread of the REB.
This is considered essential to secure scalability of the fast heating to the ignition \cite{Strozzi2012}.

\section{Magnetized fast isochoric heating}
Three critical problems have been recognized as obstacles to the efficient fast heating with the REB.
The first problem is that the REB becomes too energetic to heat the core in a long-scale-length pre-plasma filled in the cone, the second one is that a part of the REB is scattered and absorbed in a high-Z cone tip \cite{Johzaki2009a}.
The third one is that the REB has a large divergence of 100 deg. as a typical full-angle, so that only a small fraction of the diverged REB can collide with the core \cite{Bellei2013}.

The long scale-length pre-plasma filled in the cone is produced by prepulse and foot pulses of the heating laser and also by the cone breakup due to high pressure of a plasma surrounding the cone.
7\% of the heating efficiency was attained at OMEGA laser facility with an ignition-scale large area density ($\rho R \sim$ 0.3 g/cm$^2$) core \cite{Theobald2014} after reducing the pre-plasma formation in the cone \cite{Jarrott2016}, however, the aforementioned second and third critical problems still remain to be resolved.

\begin{figure*}
	\begin{center}
		\includegraphics*[width= 15cm]{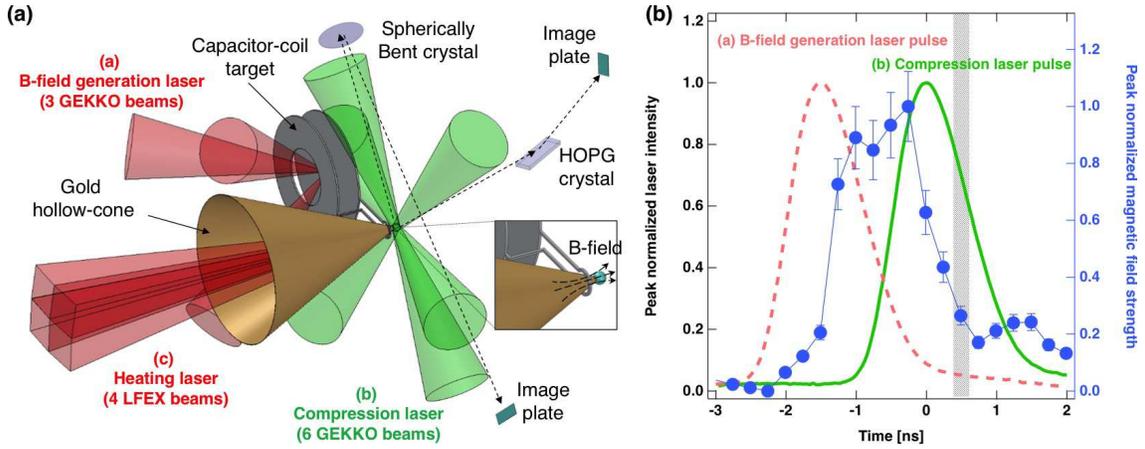}
	\end{center}
	\caption{(a) A schematic drawing of the experimental layout for the magnetized fast isochoric heating. 
(b) Timing chart of the magnetic field generation laser (pink broken line), fuel compression laser (green solid line) and laser-produced magnetic field (blue circular marks) pulses. The hatching area indicates the injection timing of the heating laser.
 \label{fig: experiment_layout}}
\end{figure*}

Here, we have introduced two novel experimental techniques.
A solid ball target is used for making an open-tip cone utilizable along with the plasma compression.
The solid ball compression does not generate preceding shocks and rarefactions travelling ahead of the dense shell, therefore a closed-tip is not required for preventing the cone inside from filling with a hot plasma.
In addition, a relatively cold and dense core can be produced stably by using the solid ball target.
The cold core enables us to visualize REB transport region \cite{Jarrott2016} in a dense core by using monochromatic Cu-$K_\alpha$ imaging technique without significant energy shift of the Cu-$K_\alpha$ photon energy due to ionization of Cu atoms.
Usage of the solid ball target has another benefit for producing a moderate guiding field as discussed later.

The other technique is a laser-driven capacitor coil target \cite{Daido1986a} to generate kilo-tesla magnetic field.
Strength of the magnetic field was measured on GEKKO-XII \cite{Fujioka2013, Law2016}, LULI2000 \cite{Santos2015a}, and OMEGA-EP laser facilities \cite{Gao2016a,Goyon2017}.
Some of the experimental results revealed that 600 - 700 T magnetic field was generated by using a tightly focused kilo-joule and nano-second infrared ($\lambda_\textrm{L}$ = 1053 nm) laser beam.

Application of external magnetic fields to the path of a REB is expected to guide the diverged REB to the dense core \cite{Strozzi2012}.
For example, the gyroradius of a 1 MeV electron under the influence of a 1 kT magnetic field is 5 $\mu$m, which is much smaller than the typical core radius (20 $\mu$m), therefore, a kilo-tesla level magnetic field is sufficient to guide the REB to the core.
The guidance of the REB by the laser-produced external magnetic field has already been demonstrated experimentally in an uncompressed-planar geometry at LULI2000 facility (Ecole Polytechnique, France) \cite{Santos2016}. 
In a more realistic configuration, the magnetic field lines are bent due to magnetic field compression associated with the plasma compression.
If a magnetic mirror is formed in the transport region, reflection of the REB by the magnetic mirror could degrade the laser-to-core coupling.
The REB transport was simulated in the mirror geometry with several mirror ratios ($R_\textrm{m}$) from 0 to 20 \cite{Johzaki2015}, and it was determined that a moderate mirror ($R_\textrm{m}$ $<$ 10) can focus the REB without significant loss caused by the mirror effect.
The magnetic field compressions were computed with the PINOCO-2D-MHD code \cite{Nagatomo2015}.
Magnetic field ratio is relatively moderate ($R_\textrm{m} \sim$ 3) in the solid ball compression compared to the gas-filled thin shell implosion because of small magnetic Reynolds number in a shock compressed solid region \cite{Fujioka2016}.

\section{Integrated experiment of magnetized fast isochoric heating}

\subsection{Laser-to-core coupling measurement}
The laser-to-core couplings were experimentally measured by varying the experimental conditions; heating laser energy, injection timing of heating laser, application of the external magnetic field or not, and open- or closed-tip cones.
The coupling was calculated from the absolute number of Cu-$K_\alpha$ X-ray (8.05 keV) photons emitted from Cu-contained pre-compressed hydrocarbon core.
Cross-sections of electron-impact $K$-shell ionization have a similar dependence on electron energy as collisional energy loss.
The two are essentially the same process but with a different threshold energy. 
Collisional deposition of REB energy (J) in a Cu-contained-core can be obtained with number of Cu-$K_\alpha$ photons (photons/sr) emitted from the core \cite{Jarrott2016} with a correlation factor.
Details of the correlation factor derivation are described in the following subsection.

Figure \ref{fig: experiment_layout} (a) shows a experimental layout.
The experiment was conducted on the GEKKO-LFEX laser facility at the Institute of Laser Engineering, Osaka University. The fusion fuel surrogate was made of a 200 μm-diameter solid Cu(II) oleate solid ball [Cu(C$_{17}$H$_{33}$COO)$_2$] \cite{Iwasa2017}, whose surface was coated with a 25 μm-thick polyvinyl alcohol (PVA) layer to prevent the Cu atoms from being ionized directly by the compression beams.
The Cu(II) oleate ball contains 9.7\% Cu atoms in weight for visualization of the relativistic electron beam (REB) transport in the core and for measurement of the laser-to-core energy coupling. 9.7\% is a little bit smaller than the ideal value (10.1\%) because of a contamination inclusion. The fuel surrogate was attached to a Au cone, whose open angle, wall thickness, and tip diameter were 45 degrees, 7 $\mu$m, and 100 $\mu$m respectively. 
The outer surface of the Au cone was coated with a 50 $\mu$m-thick PVA layer to delay the cone breakup time \cite{Fujioka2016}. Open-tip or closed-tip Au cones were used in the experiments; the tips of the closed Au cones were covered with a 7 $\mu$m-thick Au layer. 

The solid ball was compressed by six GEKKO-XII laser beams, whose wavelength, pulse shape, pulse duration, and energy were 526 nm, Gaussian, 1.3 ns full width at half maximum (FWHM), and 240 $\pm$ 15 J/beam, respectively. 
The center of the nickel-made coil, which had a 500 $\mu$m diameter, was located 230 $\mu$m from the center of the ball to apply a strong magnetic field near the REB generation point and the solid ball. The first disk of the laser-driven capacitor was irradiated through the hole of the second disk by three tightly focused GEKKO-XII laser beams, whose wavelength, pulse shape, pulse duration, and energy were 1053 nm, Gaussian, 1.3 ns (FWHM), and 600 $\pm$ 20 J/beam, respectively, yielding 7 $\times$ 10$^{15}$ W/cm$^{2}$ of intensity. 

The tip of the cone was irradiated to produce a REB by four LFEX laser beams, whose wavelength, pulse shape, and pulse duration were 1053 nm, Gaussian, 1.8 $\pm$ 0.3 ps (FWHM), respectively. The total energy of the four LFEX beams on the tip was varied from 630 to 1520 J. The focal spot diameter was 50 $\mu$m (FWHM) and contained 30\% of the total energy within the 50 $\mu$m-diameter, producing an intensity of 1.3 $\times$ 10$^{19}$ W/cm$^2$ at the maximum energy shot.

Copper $K_\alpha$ X-rays (8.05 keV) were imaged using a spherically bent quartz (2131) crystal to visualize the transport of the REB in a pre-compressed core from the direction perpendicular to the LFEX incident axis. The magnification, spatial resolution, and spectral bandwidth were 20, 13 $\mu$m (FWHM), and 5 eV (FWHM), respectively.

The X-ray spectrometer, which utilizes a planar highly oriented pyrolytic graphite (HOPG), was installed at 40 deg. from the LFEX incident axis to measure the absolute Cu-$K_\alpha$ yield. The absolute integral reflectance of the HOPG was measured using an X-ray diffractometer with an accuracy of $\pm$15\%. 
Spatial non-uniformity of the integral reflectance of the HOPG is the dominant source of this error. The spectral resolution of the spectrometer was 17.9 eV (FWHM). 

\begin{figure}
	\begin{center}
		\includegraphics*[width= 8cm]{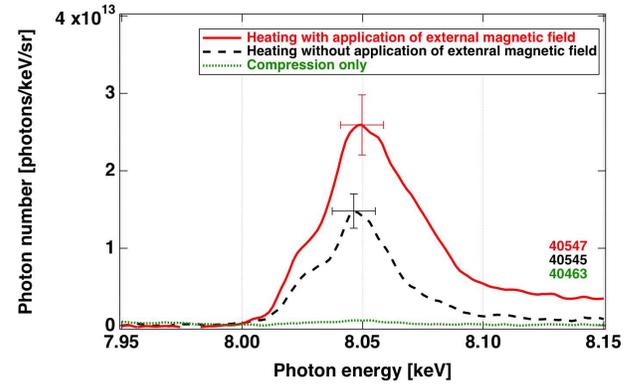}
	\end{center}
	\caption{Example Cu-$K_\alpha$ spectra peaked at 8.05 keV. Red solid, black dashed, and green dotted lines are, respectively, spectra obtained by heating with application of external magnetic field, heating without the application, and only fuel compression. The vertical error bar corresponds to the 15\% error in the absolute integrated reflectance of the HOPG. The horizontal error bar is equal to 17.9 eV of the spectral resolution of the spectrometer. The Cu-$K_\alpha$ photon yields were integrated within the energy range of 8.0 keV to 8.1 keV.
 \label{fig: x-ray_spectra}}
\end{figure}

Figure \ref{fig: x-ray_spectra} shows example Cu-$K_\alpha$ spectra. 
Cu-$K_\alpha$ X-ray yield produced during the compression process (green dotted line) was negligibly weak compared to those produced during the heating process (red solid and black dashed lines).  The vertical error bar corresponds to the error in the integral reflectance measurement, and the horizontal error bar is equal to the spectral resolution. The Cu-Kα photon yields were integrated within the energy range of 8.0 keV to 8.1 keV after background subtraction. 

The laser energy and injection timing of the heating laser are summarized in Table \ref{table: shot_summary}.
The injection timings were measured with an x-ray streak camera with an accuracy of $\pm$ 0.02 ns.
Figure \ref{fig: experiment_layout} (b) shows a time chart of the magnetic-field-generation laser, compression laser and laser-produced magnetic field pulses. The time origin ($t$ = 0 ns) is defined as the peak of the compression laser pulse.
The peak of the magnetic field generation laser pulse was set at $t$ = -1.5 ns, therefore the magnetic field strength reaches its maximum value before the compression beam irradiation.
The heating lasers (four LFEX beams) were injected around the maximum compression timing $t$ = 0.38 - 0.72 ns shown as the hatching area in Fig. \ref{fig: experiment_layout} (b).

\subsection{Initial magnetic field profile calculation}
An externally applied magnetic field penetrates diffusively into the target from the outside. The diffusion time scale is determined by the electrical conductivity and spatial size of the targets. Based on a previous study, the externally applied magnetic field is guaranteed to penetrate rapidly into an insulator hydrocarbon\cite{Santos2016}, however, the diffusion dynamics of the magnetic field into the gold cone remain unclear. 

The magnetic diffusion time ($t_\textrm{diff}$) is expressed as $t_\textrm{diff}$ = $\mu_0 \sigma L^2$, where $\mu_{0}$, $\sigma$ and $L$ are the permeability, electrical conductivity, and diffusion layer thickness, respectively.
The temporal change in magnetic field strength drives an eddy current in the gold cone, and the current ohmically heats the gold. The electron conductivity of the gold depends on its temperature.
For a 7 $\mu$m gold cone wall, the diffusion times are $t_\textrm{diff}$ = 2.5 ns, 120 ps and 60 ps at room temperature ($\sigma$ = 4 $\times$ 10$^7$ S/m), 0.1 eV ($\sigma$ = 2 $\times$ 10$^6$ S/m) and 1 eV ($\sigma$ = 1 $\times$ 10$^6$ S/m), respectively \cite{Dharma-wardana2006}.
The small temperature increment helps to rapidly soak the cone in the magnetic field.

250 kA of current was generated with a capacitor-coil target driven by one GEKKO-XII beam, which was measured using proton radiography \cite{Law2016}.
430 kA of current, which is $\sqrt{3}$ times larger than that generated with one beam, can be driven by using three GEKKO-XII beams.

\begin{figure}
	\begin{center}
		\includegraphics*[width= 8cm]{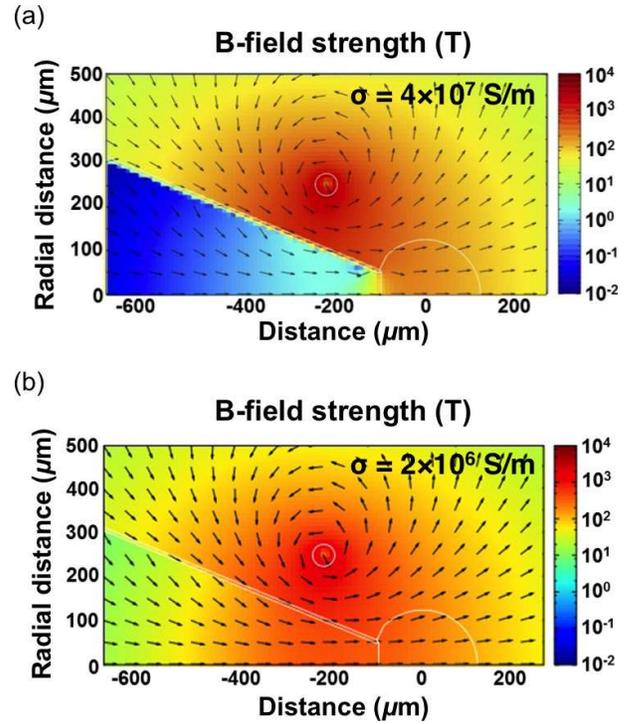}
	\end{center}
	\caption{Two-dimensional profile of the magnetic field generated with a coil, in which 430 kA of current flows at the field peak timing. The magnetic field diffusion was taken into account using two electrical conductivities [(a)  = 4 $\times$ 10$^7$ S/m and (b) 2 $\times$ 10$^6$ S/m]. 
 \label{fig: B-field_profile}}
\end{figure}

Figure \ref{fig: B-field_profile} shows the two-dimensional profile of the magnetic field calculated at the maximum magnetic field strength timing generated using a 430 kA Gaussian current pulse with a 1 ns (FWHM) pulse width. The diffused magnetic field profiles were calculated using two different electrical conductivities ($\sigma$ = 4 $\times$ 10$^7$ and 2 $\times$10$^6$ S/m) and neglecting temporal changes in the temperature, density, and conductivity of the gold cone.
The magnetic field strengths around the tip are 320 and 430 T, respectively, in this calculation.
These strengths are high enough to guide the REB.

\subsection{Two-dimensional density profile measurement}
Flash X-ray backlighting with a monochromatic imager \cite{King2005, Fujioka2010, Theobald2014, Sawada2016} was used to measure two-dimensional density profiles of pre-compressed Cu(II) oleate solid balls under an external magnetic field. 

\begin{figure*}
	\begin{center}
		\includegraphics*[width= 15cm]{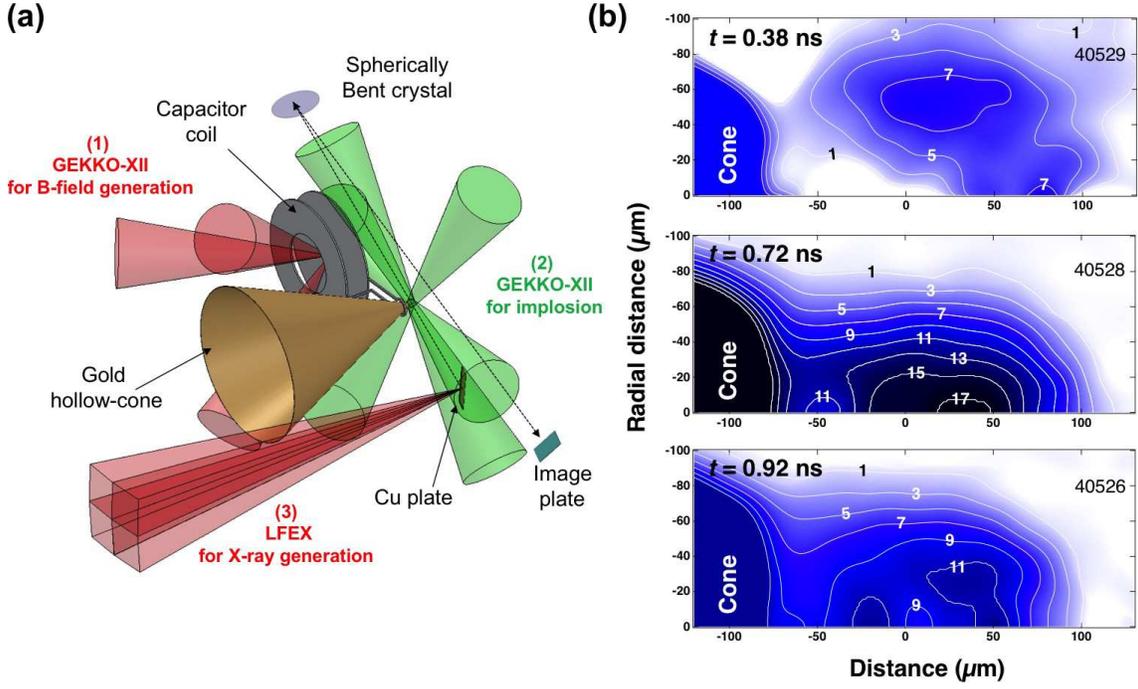}
	\end{center}
	\caption{(a) Experimental layout of the pre-compressed core density measurement experiment. The LFEX laser was used to generate a Cu-$K_\alpha$ backlight flash. (b) Density profiles measured at $t$ = 0.38, 0.72, and 0.92 ns after the peak of the compression beam pulse.
 \label{fig: density_measurement}}
\end{figure*}

The experimental layout is shown in Fig. \ref{fig: density_measurement}(a).
The X-ray shadows of compressed cores were imaged using imaging plates with the same spherically bent quartz crystal used in the laser-to-core coupling experiment.
The solid ball specifications and the laser parameters of the compression and magnetic field generation beams were also identical to those used in the laser-to-core coupling measurement. The LFEX laser was used for flash Cu-$K_\alpha$ X-ray backlight generation in this experiment. The LFEX laser was defocused to produce a 350 $\mu$m-diameter spot on a 20 $\mu$m-thick Cu foil at 3 mm behind the solid ball along the line of sight of the crystal imager to generate a large format backlight.

An X-ray shadow is converted to an X-ray transmittance profile by interpolating the two-dimensional backlight X-ray intensity profile within the core region from the outside of the core region. The area density of the pre-compressed core was calculated from the X-ray transmittance profile with a calculated opacity of 100 eV Cu(II) oleate for 8.05 keV X rays \cite{Macfarlane2006a}. 
A two-dimensional density profile of the core was obtained after applying an inverse Abel transformation to the area density profile, assuming rotational symmetry of the core along the cone axis. 

Figure \ref{fig: density_measurement}(b) shows the core density profiles at $t$ = +0.38, +0.72, and +0.92 ns. The converging shock wave was still travelling to the center of the ball at $t$ = +0.38 ns, maximum compression was reached at around $t$ = +0.72 ns, and the core had already begun to disassemble at $t$ = +0.92 ns.
The area mass densities ($\rho L$) and average mass density ($\rho$) of the core along the REB path length ($L$) were, respectively, $\rho L$ = 0.08 g/cm$^2$ and $\rho$ = 5.7 g/cm$^3$ at $t$ = +0.38 ns, $\rho L$ = 0.16 g/cm$^2$ and $\rho$ = 11.3 g/cm$^3$ at $t$ = +0.72 ns.
These values were used in the calculation of the correlation factor described in the next section.

\subsection{Derivation of correlation factor between Cu-$K_\alpha$ photons and deposited REB energy}

The Solodov model \cite{Solodov2008} was used to calculate the stopping power $S(E, \rho)$ of the REB in a core, here $E$ is electron kinetic energy.
Both Davies \cite{Davies2013} and Hombourger \cite{Hombourger1998} models were used to calculate the electron-impact K-shell ionization cross-section $\sigma_{K_\alpha} (E)$. 
The differences in cross-section between the two models are considered to be model-dependent errors. The ratio between the stopping power and $K$-shell ionization cross section is the correlation factor ($C$). Note that the Hombourger model gives a 1.3 times higher correlation factor than the Davies model. The correlation factor $C$ is defined as follows,

\begin{eqnarray}
C & = & \frac{\int^{L}_{0} \int^{\textrm{1 GeV}}_{\textrm{10 keV}} \varepsilon_\textrm{dep} (E, x) dE dx}{\int^{L}_{0} \int^{\textrm{1 GeV}}_{\textrm{10 keV}} P_{K_\alpha} (E, x) dE dx} \\
& = & \frac{\int^{L}_{0} \int^{\textrm{1 GeV}}_{\textrm{10 keV}} v(E, x) f(E, x) S(E, \rho) dE dx}{\int^{L}_{0} \int^{\textrm{1 GeV}}_{\textrm{10 keV}} v(E, x) f(E, x) \sigma_{K_\alpha}(E)/4\pi dE dx} 
\end{eqnarray}

where $\varepsilon_\textrm{dep} (E, x)$, $P_{K \alpha} (E, x$), $v(E, x)$, $f(E, x)$, and $n_\textrm{Cu}$ are the collisionally deposited energy by the REB to the dense core, the probability of Cu-Kα emission, the relativistic electron velocity, the REB energy distribution, and the number density of Cu atoms in a core, respectively.
The initial REB energy distribution was a Boltzmann function as $f(E, 0) = \exp (- E/T_\textrm{REB})$ at the generation point ($x$ = 0), here TREB is the slope temperature of the energy distribution. Slowing down of the REB during the transport was considered.

\begin{figure}
	\begin{center}
		\includegraphics*[width= 8cm]{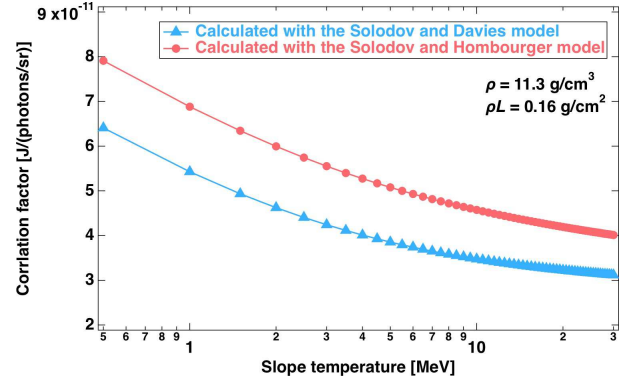}
	\end{center}
	\caption{Dependence of the correlation factor between the deposited energy (J) and Cu-$K_\alpha$ yield (photons/sr) on the REB slope temperature. The data were calculated using the Davies (blue circular marks) and Hombourger (red triangular marks) models of electron-impact K-shell ionization for a 11.3 g/cm$^3$ and 0.16 g/cm$^2$ Cu(II) oleate, which correspond to the average core density and area density at $t$ = 0.72 ns (maximum compression timing). .
 \label{fig: corr_factor}}
\end{figure}

Figure. \ref{fig: corr_factor} shows the calculated correlation factor at different REB slope temperatures ($T_\textrm{REB}$) for a $\rho$ = 11.3 g/cm$^3$ and $\rho L$ = 0.16 g/cm$^2$ Cu(II) oleate plasma, which are equal to those observed at $t$ = +0.72 ns.
The correlation factor also depends weakly on the core density. Dependence of the correlation factor on $T_\textrm{REB}$ must be considered in the coupling evaluation.

Energy distributions of electrons that escaped from the plasma into vacuum were measured using an electron energy analyzer positioned along the LFEX incident axis.
Although the energy distribution of the so-called vacuum electrons is not exactly identical to that at the generation point due to scattering, absorption, and reflection by the cone, core, and spontaneous electromagnetic field, we found that the slope temperatures of the escaped electrons are close to those in the transport region estimated from Bremsstrahlung X-ray spectra \cite{Fujioka2015}.
Therefore, the energy distribution of the measured vacuum electrons was used in the correlation factor calculation. The energy distribution of the escaped electrons was fitted with a two-temperature Boltzmann distribution function as $f(E) = A \exp (- E/T_{\textrm{REB1}}) + (1 - A) \exp (- E/T_{\textrm{REB2}})$, where $A$ and $E$ are the intercept and electron energy, and $T_\textrm{REB1}$ $<$ $T_\textrm{REB2}$. 

The correlation factor calculated for each shot is summarized in Table \ref{table: corr_factor} along with other measured parameters.
The error in the deposited energy was evaluated taking into account the difference in the correlation factor between the ionization models and the error in the integral reflectance of the HOPG.

\begin{table*}
	\caption{\label{table: corr_factor} Summary of correlation factors used in the analysis}
	\begin{tabular}{cccccc}
		\hline
		\textbf{Shot} & \textbf{$A$} & \textbf{$T_\textrm{REB1}$} & \textbf{$T_\textrm{REB2}$} & \textbf{Correlation factor} & \textbf{Correlation factor}\\
		\textbf{ID} & & [MeV] & [MeV] & \textbf{with Davis model} & \textbf{with Hombourger model}\\
 		& & & & [J/photons/str] & [J/photons/str] \\\hline\hline
		40545 & 0.881 & 1.0 & 4.7 & 4.1 $\times$ 10$^{-11}$ & 5.3 $\times$ 10$^{-11}$\\
		40541 & 0.951 & 0.7 & 4.4 & 4.2 $\times$ 10$^{-11}$ & 5.5 $\times$ 10$^{-11}$\\\hline
		40558 & 0.956 & 4.6 & 23.6 & 3.4 $\times$ 10$^{-11}$ & 4.5 $\times$ 10$^{-11}$\\
		40556 & 0.933 & 2.2 & 5.4 & 3.7 $\times$ 10$^{-11}$ & 4.9 $\times$ 10$^{-11}$\\
		40547 & 0.907 & 1.6 & 2.8 & 4.1 $\times$ 10$^{-11}$ & 5.4 $\times$ 10$^{-11}$\\
		40549 & 0.999 & 0.8 & 10 & 4.7 $\times$ 10$^{-11}$ & 6.0 $\times$ 10$^{-11}$\\
		40543 & 0.971 & 0.5 & 4.1 & 4.5 $\times$ 10$^{-11}$ & 5.8 $\times$ 10$^{-11}$\\\hline
		40560 & 0.991 & 0.9 & 21.7 & 3.9 $\times$ 10$^{-11}$ & 5.2 $\times$ 10$^{-11}$\\
		40562 & 0.890 & 1.5 & 5.6 & 4.0 $\times$ 10$^{-11}$ & 5.2 $\times$ 10$^{-11}$\\\hline
	\end{tabular}
\end{table*}

Table \ref{table: shot_summary} summarizes laser-to-core coupling efficiency obtained in this experiment.
The data are categorized into three groups according to the experimental conditions, where the external magnetic field was applied or not, and the cone-tip was open or closed.
The data are sorted by the laser-to-core coupling efficiency in each group.

\begin{table*}
	\caption{\label{table: shot_summary} Summary of laser-to-core coupling efficiencies}
	\begin{tabular}{cccccccc}
		\hline
		\textbf{Shot} & \textbf{Cone} & \textbf{Heating} & \textbf{Compression} & \textbf{B-generation} & \textbf{Heating} & \textbf{Cu-K$\alpha$} & \textbf{Coupling} \\
		\textbf{ID} & \textbf{Tip} & \textbf{Energy} & \textbf{Energy} & \textbf{Energy} & \textbf{Timing} & \textbf{number} & \textbf{Efficiency} \\
		 & \textbf{Condition} & [J] & [J] & [J] & [ns] & [photons/sr] & [\%] \\\hline \hline
		40545 & Open & 899 & 1422 & \textbf{N/A} & 0.42 & 5.58 $\times$ 10$^{11}$ & 2.9 $\pm$ 0.5 \\
		40541 & Open & 683 & 1428 & \textbf{N/A} & 0.65 & 5.53 $\times$ 10$^{11}$ & 3.9 $\pm$ 0.7 \\\hline
		40558 & Open & 1516 & 1386 & 1761 & 0.4 & 1.19 $\times$ 10$^{12}$ & 3.1 $\pm$ 0.5 \\
		40556 & Open & 1016 & 1332 & 1698 & 0.61 & 1.02 $\times$ 10$^{12}$ & 4.3 $\pm$ 0.8 \\
		40547 & Open & 1100 & 1530 & 1824 & 0.38 & 1.28 $\times$ 10$^{12}$ & 5.5 $\pm$ 1.0 \\
		40549 & Open & 668 & 1548 & 1794 & 0.37 & 7.29 $\times$ 10$^{11}$ & 5.8 $\pm$ 1.0 \\
		40543 & Open & 625 & 1494 & 1842 & 0.72 & 9.32 $\times$ 10$^{11}$  & 7.7 $\pm$ 1.3 \\\hline
		40560 & \textbf{Close} & 1523 & 1404 & 1794 & 0.38 & 8.23 $\times$ 10$^{11}$  & 2.5 $\pm$ 0.4\\
		40562 & \textbf{Close} & 1378 & 1374 & 1725 & 0.65 & 7.96 $\times$ 10$^{11}$  & 2.7 $\pm$ 0.5\\\hline
	\end{tabular}
\end{table*}

\section{Discussion}
\begin{figure*}
	\begin{center}
		\includegraphics*[width= 15cm]{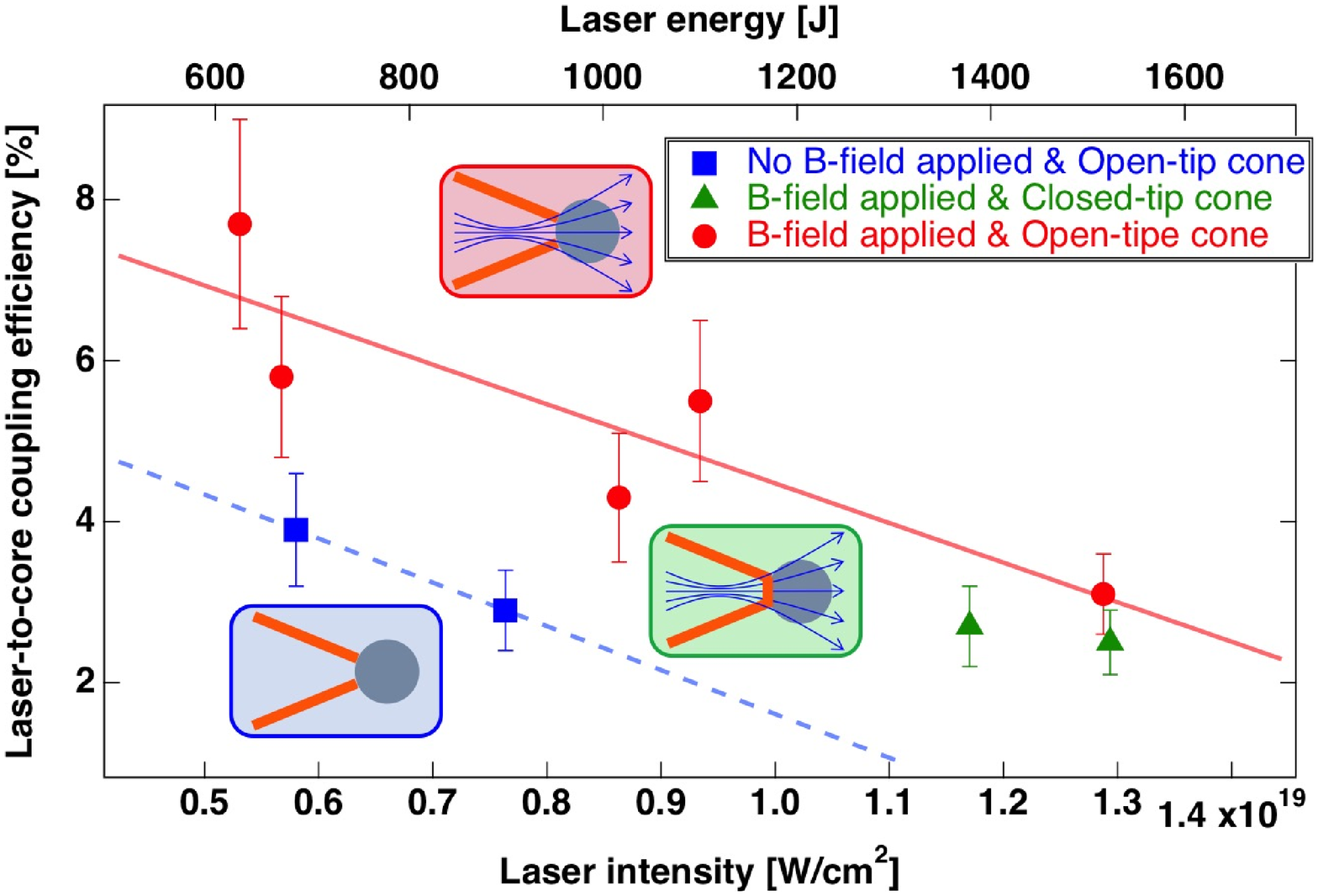}
	\end{center}
	\caption{Dependence of laser-to-core energy coupling on heating laser intensity (bottom axis) and energy (top axis). The blue rectangular, green triangle and red circle marks represent laser-to-core coupling efficiencies obtained with the following conditions; no application of external-magnetic-field with open-tip cone, application of external magnetic field with closed-tip cone, and application of external magnetic field with open-tip cone, respectively.
 \label{fig: laser-to-core-coupling}}
\end{figure*}

Figure \ref{fig: laser-to-core-coupling} shows the dependence of the measured coupling efficiency on the heating laser intensity (bottom axis) and energy (top axis).
The figure reveals clearly that the REB guiding by the external magnetic field (red circular marks) shows 1.8 times higher coupling than that without the guiding (blue rectangular marks), and that electron scattering in the closed cone-tip (green triangular marks) reduces the coupling 0.8 times compared to the open-tip values (red circular marks).
Furthermore, the coupling was degraded gradually by increasing the heating laser energy with keeping both pulse duration and spot diameter unchanged (1.8 ps and 50 $\mu$m), because a higher intensity laser produces higher temperature REB and this results in less coupling.
The pulse duration must be extended to sustain this efficient coupling for higher laser energy.

\begin{figure*}
	\begin{center}
		\includegraphics*[width= 15cm]{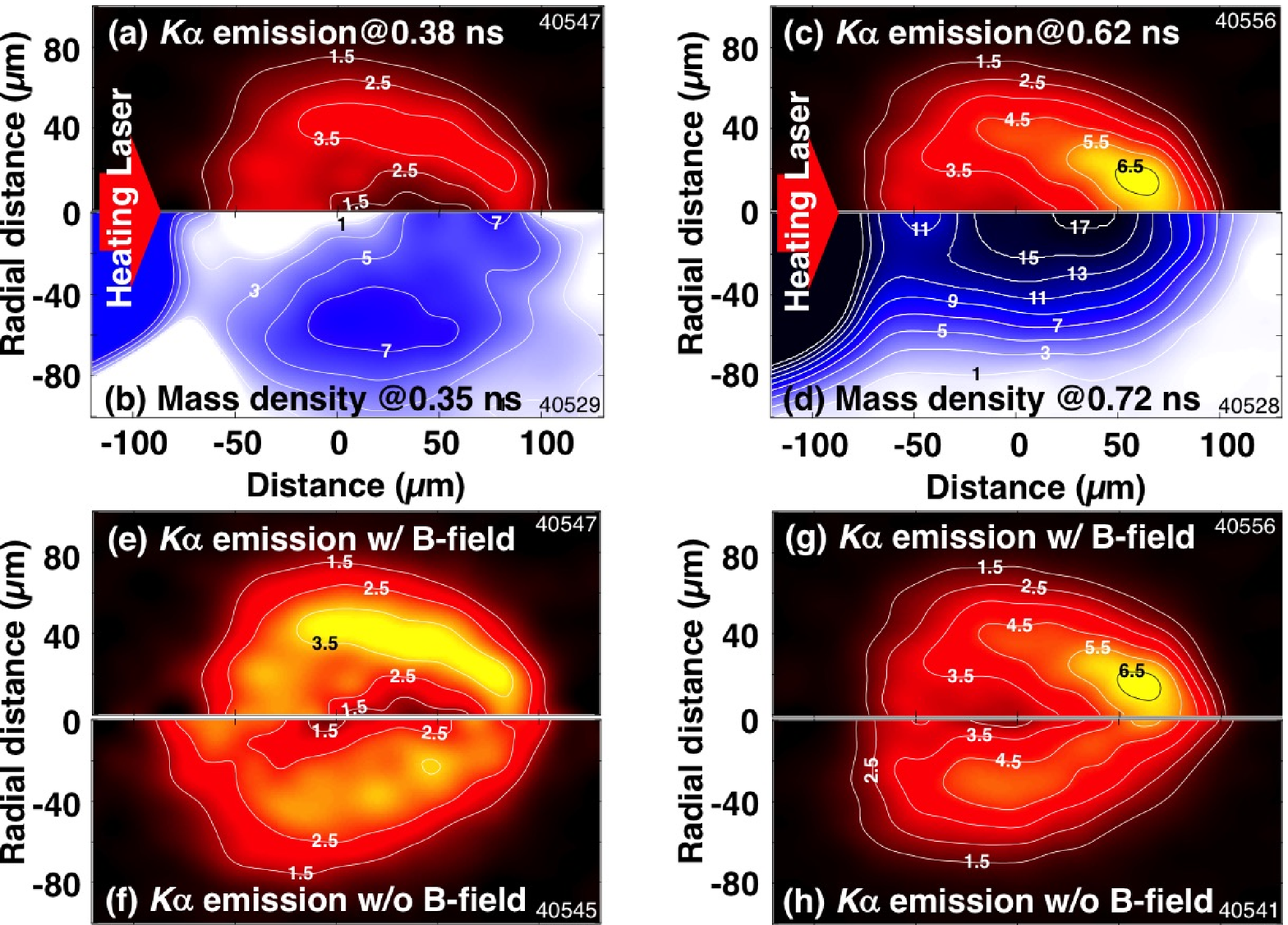}
	\end{center}
	\caption{Two dimensional profiles of Cu-$K_\alpha$ emission (a,c,e,g) and mass density (b,d) measured in the experiments with the application of the external magnetic field at $t$ = 0.40 $\pm$ 0.03 ns (a, b, e) and $t$ = 0.67 $\pm$ 0.05 ns (c,d,g). 
Cu-$K_\alpha$ emission profiles are compared between those obtained with (e,f) and without (g,h) application of the external magnetic field at two different injection timings.
The Cu-$K_\alpha$ emission profiles (a, c, g, h) are drawn with the same color scale excepting for (e) and (f).
These images were obtained after applying an inverse Abel transformation to the area density profile, assuming rotational symmetry of the core along the cone axis.
\label{fig: density_ka_comparison}}
\end{figure*}

Figure \ref{fig: density_ka_comparison} shows two dimensional Cu-$K_\alpha$ emission profiles and density profiles of the pre-compressed core at two different timings ($t$ = 0.40 $\pm$ 0.03 ns and 0.67 $\pm$ 0.05 ns) and also the comparison of Cu-$K_\alpha$ emission profiles between with and without the external magnetic field application at the two different timings.
Here the density profiles were measured in separate experiments that are explained in the Supplemental Information.

At $t$ = 0.40 $\pm$ 0.03 ns, the converging shock wave was still travelling to the center of the ball, and the shock front is clearly observable in Fig. \ref{fig: density_ka_comparison} (b).
Strong Cu-$K_\alpha$ emission region locates near the shock front  in Fig. \ref{fig: density_ka_comparison} (a, e).
Externally applied magnetic field lines were accumulated in the downstream of the shock wave, therefore REB was guided to the shock compressed region along the field lines.
This feature was not observed without the external magnetic field as shown in Fig. \ref{fig: density_ka_comparison} (f).

At $t$ = 0.67 $\pm$ 0.05 ns, the solid ball reached the maximum compression.
The strong Cu-$K_\alpha$ emission spot appeared at 50 $\mu$m longitudinal distance, which was more than 100 $\mu$m away from the REB generation point namely the cone tip as shown in Fig. \ref{fig: density_ka_comparison} (c, g).
This Cu-$K_\alpha$ emission feature is the evidence of the REB guiding by the externally applied magnetic field as well as significant enhancement of the laser-to-core coupling.
This strong emission spot disappeared, when the external magnetic field was not applied as shown in Fig. \ref{fig: density_ka_comparison} (h).
Energy shift of the Cu-$K_\alpha$ X ray due to ionization of Cu atoms in a hot core could be the reason why the Cu-$K_\alpha$ emission was weak in the core central region \cite{Akli2007}.

The laser-to-core coupling ($\eta$) can be simplified as a product of laser-to-REB energy conversion efficiency ($\eta_\textrm{REB}$), REB collision probability ($\eta_\textrm{coll}$), and energy deposition rate of REB in the core ($\eta_\textrm{dep}$) \cite{Fujioka2015}.

\begin{equation}
\eta  = \eta_\text{REB} \cdot \eta_\text{col} \cdot \eta_\text{dep} \\
\end{equation} 

The previous experiments show $\eta_\textrm{REB}$ = 0.4 \cite{Fujioka2015} and $\eta_\textrm{coll}$ = 0.7 \cite{Santos2016}. 
$\eta_\textrm{dep}$ was calculated by the simplified model \cite{Fujioka2015} with the measured area density and REB temperatures shown in Table \ref{table: corr_factor}

\begin{widetext}
\begin{equation}
\eta_\text{dep}
= \frac{A T^2_\text{REB1}}{AT^2_\text{REB1} + (1-A)T^2_\text{REB2}} \cdot \frac{\rho L}{0.6 T_\text{REB1}}+\frac{(1-A) T^2_\text{REB2}}{AT^2_\text{REB1} + (1-A)T^2_\text{REB2}} \cdot \frac{ \rho L}{0.6 T_\text{REB2}}.
\end{equation}
\end{widetext}

An approximated relation of Eq. (11) in Ref. \cite{Atzeni2005}, $R_\text{REB}$ [g/cm$^2$] = 0.6 $f_\text{R}$ $T_\text{REB}$ [MeV], is used to calculate the $R_\text{REB}$ from the experimentally measurable parameter ($T_\text{REB}$), here $f_\text{R}$ is an adjustable parameter, set to 1 in the standard model.
Finally, the simple model yields $\eta = \eta_\textrm{REB} \cdot \eta_\textrm{coll} \cdot \eta_\textrm{dep}$ = 6.2\%.
This simple evaluation seems consistent fairly with the measured coupling (7.7 $\pm$ 1.4 \%), and the simple evaluation reveals that higher area density core leads to higher laser-to-core coupling.
An ultra-high-energy density state could be efficiently created by the magnetized fast isochoric heating.

\section{Conclusion}
We have studied experimentally the magnetically-reinforced fast isochoric heating scheme for creating ultra-high-energy-density state related to the ignition spark formation.
The enhancement of the laser-to-core coupling as well as the strong Cu-$K_\alpha$ emission spot located at 100 $\mu$m away from the cone tip are the evident features produced by guiding of the diverged REB with the externally applied magnetic field in the long transport distance.
An energy density increment of the heated core is close to 1 Gbar, which corresponds to 50 J of the energy deposition in a 100 $\mu$m-diameter spherical volume.
Plasma hydrodynamics, generation and transport of electron/ion beams, thermal conduction and $\alpha$ particle transport will be able to be controlled by the externally applied strong magnetic field.
There is no doubt that laser-plasma experiments with strong magnetic fields contain a lot of unexplored physics, therefore this research also stimulates spin-off sciences in the field of atomic physics, nuclear physics, and astrophysics which act to broaden inertial fusion sciences and high energy density sciences.

\begin{acknowledgments}
The authors thank the technical support staff of ILE and the Cyber Media Center at Osaka University for assistance with the laser operation, target fabrication, plasma diagnostics, and computer simulations.
The authors appreciate greatly valuable discussions with Drs. J. Moody, B. Pollock, M. Tabak, W. Kruer, O. Landen, T. Ma, H. Chen, A. Kemp, D. Mariskal and B. Remington (LLNL), Prof. M. Murakami, Drs. K. Shigemori and T. Sano (ILE, OU), Dr. Iwamoto (NIFS), and especially Dr. S. Wilks (LLNL) also for his proof-reading of this manuscript.
This work was supported by the Collaboration Research Program between the National Institute for Fusion Science and the Institute of Laser Engineering at Osaka University, and by the Japanese Ministry of Education, Science, Sports, and Culture through Grants-in-Aid, KAKENHI (Grants No. 24684044, 25630419, 15K17798, 15K21767, 15KK0163, 16K13918, 16H02245, and 17K05728), Bilateral Program for Supporting International Joint Research by JSPS, and Grants-in-Aid for Fellows by Japan Society for The Promotion of Science (Grant No. 14J06592, 15J00850, and 15J02622).
The study also benefited from diagnostic support funded by the French state through research projects TERRE ANR-2011-BS04-014 (French National Agency for Research (ANR) and Competitiveness Cluster "Route des Lasers") and ARIEL (Regional Council of Aquitaine). M. B.-G. and J. J. S. acknowledge the financial support received from the French state and managed by ANR in the framework of the "Investments For the Future" program at IdEx Bordeaux - LAPHIA (ANR-10-IDEX-03-02), from COST Action MP1208 "Developing the Physics and the Scientific Community for Inertial Fusion" and from the Euratom research and training program 2017-2018 under grant agreement "StarkZee" No CfP-AWP17-IFE-CEA-02.
The views and opinions expressed herein do not necessarily reflect those of the European Commission.
\end{acknowledgments}


\end{document}